\documentclass[10pt]{iopart}

\usepackage{graphicx}% Include figure files
\usepackage{iopams}  
\begin{document}

\title[Fate of Ising ferromagnets and antiferromagnets]{Fate of Ising ferromagnets and antiferromagnets by zero-temperature Glauber dynamics on the two-dimensional 
Archimedean and 2-uniform lattices}

\author{Unjong Yu}
\address{Department of Physics and Photon Science,
              Gwangju Institute of Science and Technology, 
              Gwangju 61005, Republic of Korea}
\ead{uyu@gist.ac.kr}

\date{\today}

\begin{abstract}
The fate of the Ising ferromagnet and antiferromagnet by the zero-temperature Glauber dynamics
from random initial spin configuration is investigated
in the two-dimensional Archimedean and 2-uniform lattices.
Blinker states are found in addition to the ground state and metastable state.
We show that an even-coordinated lattice can arrive at a
blinker state or a metastable state without stripe structure,
in contrast to common expectation.
The universal relationship between the critical percolation
and the probability of stripe final state
is confirmed for six lattices.
Results about the fate of the antiferromagnetic Ising model show that the geometric frustration 
suppresses ordering more and promotes blinker state.
\end{abstract}

%\pacs{05.20.-y, 05.70.Ln, 64.60.De, 05.50.+q}
%05.20.-y	Classical statistical mechanics
%05.70.Ln	Non-equilibrium and irreversible thermodynamics
%64.60.De	Statistical mechanics of model systems (Ising model, Potts model, field-theory models, Monte Carlo techniques, etc.)
%05.50.+q	Lattice theory and statistics (Ising, Potts, etc.)

 Frustrated systems classical and quantum

%\maketitle

\section{Introduction}

%%%%%%%%%%%%%%%%%%%%%%%%%%%%%%
\begin{figure}[tb!]
\includegraphics[width=13.0cm]{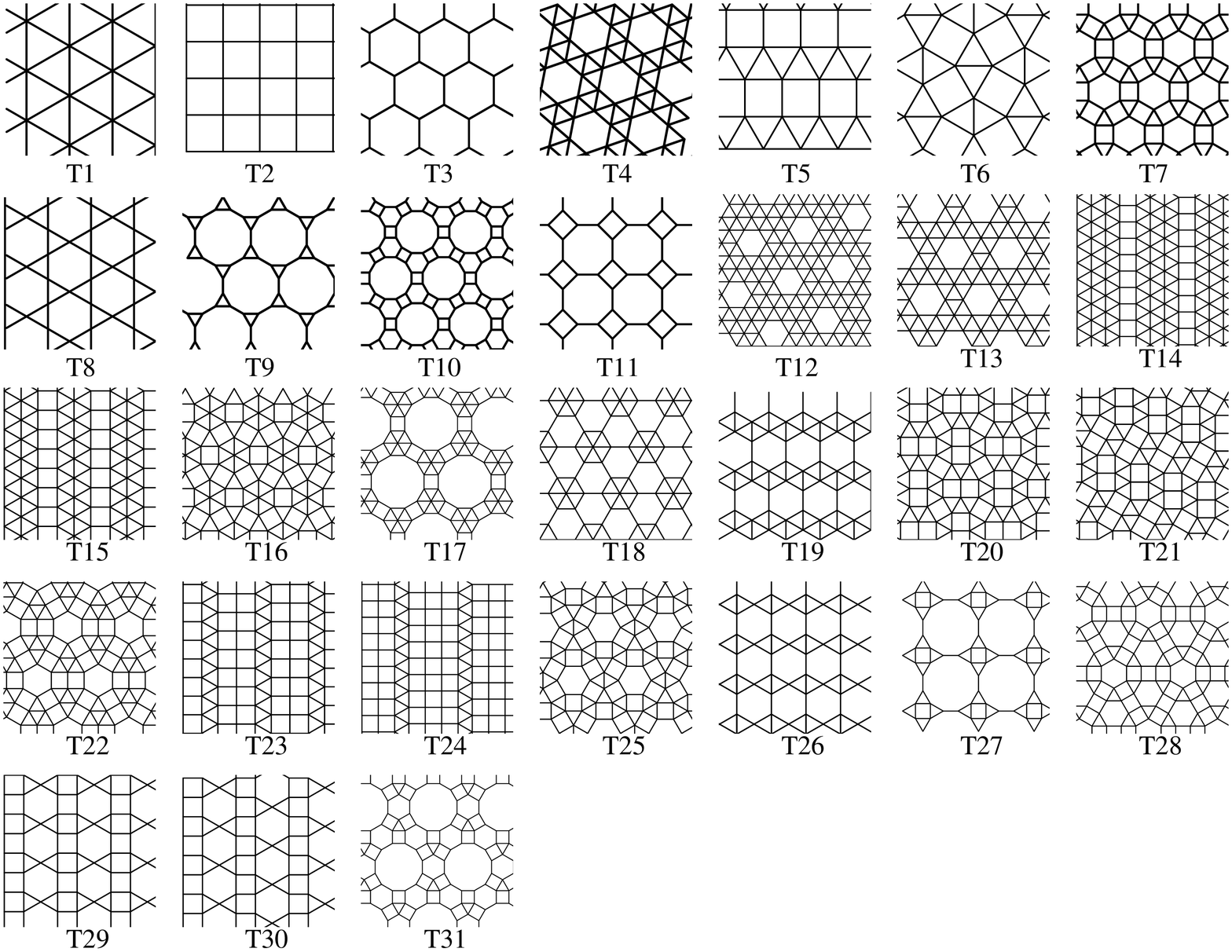}
\caption{The Archimedean lattices (T1$\sim$T11) and
the 2-uniform lattices (T12$\sim$T31).}
\label{Fig_Arch}
\end{figure}

%\begin{table}[tb!]
\Table{\label{Table_fate} Name, coordination number ($z_1$ and $z_2$),
and fate of the zero-temperature Glauber dynamics
with ferromagnetic (F) and antiferromagnetic (AF) nearest-neighbor interaction
in the Archimedean and 2-uniform lattices.
For the Archimedean lattices, $z_2$ is meaningless
because there is only one kind of lattice point.
In the two rightmost columns, G, M, and B represent
the long-range ordered ground state, metastable state,
and blinker state, respectively.
}
%\begin{tabular}{ccccccc}
\br
\multicolumn{2}{c}{Name} & $z_1$ & $z_2$ & Fate(F) & Fate(AF)\\
\mr
T2  & ($4^4$)                        & 4 & - &  G+M & - \\
T1  & ($3^6$)                        & 6 & - &  G+M & B\\
T8  & ($3,6,3,6$)                    & 4 & - &  G+B & B\\
T30 & ($3, 4^2, 6; 3, 6, 3, 6$)$_2$  & 4 & 4 &  G+B & B$^{\rm a}$ \\
T26 & ($3^2, 6^2; 3, 6, 3, 6$)       & 4 & 4 &  G+M+B & G+M\\
T29 & ($3, 4^2, 6; 3, 6, 3, 6$)$_1$  & 4 & 4 &  G+M+B & B \\
\hline
T3  & ($6^3$)                        & 3 & - &  M & - \\
T10 & ($4,6,12$)                     & 3 & - &  M & - \\
T11 & ($4,8^2$)                      & 3 & - &  M & - \\
T4  & ($3^4,6$)                      & 5 & - &  M & M \\
T5  & ($3^3,4^2$)                    & 5 & - &  M & M \\
T6  & ($3^2,4,3,4$)                  & 5 & - &  M & M \\
T9  & ($3,12^2$)                     & 3 & - &  M & M \\
T20 & ($3^3, 4^2; 3^2, 4, 3, 4$)$_1$ & 5 & 5 &  M & M \\
T21 & ($3^3, 4^2; 3^2, 4, 3, 4$)$_2$ & 5 & 5 &  M & M \\
T7  & ($3,4,6,4$)                    & 4 & - &  M & B \\
T13 & ($3^6; 3^4, 6$)$_2$            & 5 & 6 &  M & B \\
T17 & ($3^6; 3^2, 4, 12$)            & 4 & 6 &  M & B \\
T18 & ($3^6; 3^2, 6^2$)              & 4 & 6 &  M & B \\
T19 & ($3^4, 6; 3^2, 6^2$)           & 4 & 5 &  M & B \\
T27 & ($3, 4, 3, 12; 3, 12^2$)       & 3 & 4 &  M & B \\
T31 & ($3, 4, 6, 4; 4, 6, 12$)       & 3 & 4 &  M & B \\
\hline
T12 & ($3^6; 3^4, 6$)$_1$            & 5 & 6 &  B & B \\
T14 & ($3^6; 3^3, 4^2$)$_1$          & 5 & 6 &  B & B \\
T15 & ($3^6; 3^3, 4^2$)$_2$          & 5 & 6 &  B & B \\
T16 & ($3^6; 3^2, 4, 3, 4$)          & 5 & 6 &  B & B \\
T22 & ($3^3, 4^2; 3, 4, 6, 4$)       & 4 & 5 &  B & B \\
T23 & ($3^3, 4^2; 4^4$)$_1$          & 4 & 5 &  B & B \\
T24 & ($3^3, 4^2; 4^4$)$_2$          & 4 & 5 &  B & B \\
T25 & ($3^2, 4, 3, 4; 3, 4, 6, 4$)   & 4 & 5 &  B & B \\
T28 & ($3, 4^2, 6; 3, 4, 6, 4$)      & 4 & 4 &  B & B \\
\br
\end{tabular}
\item[] $^{\rm a}$ Blinker state with partial one-directional long-range ordering.
\end{indented}
\end{table}
%%%%%%%%%%%%%%%%%%%%%%%%%%%%%%

The ground state of the ferromagnetic Ising model \cite{Ising25} is simply ferromagnetic phase
with all spins in the same direction with one another.
It is not trivial, however, when the system
is quenched abruptly from a high-temperature disordered phase to
below the critical temperature. Especially, quenching to zero-temperature is
unique because the system trapped by any finite energy barrier cannot escape
from the metastable state forever.

When the system is quenched from infinite temperature (random spin configuration)
to zero temperature by the zero-temperature Glauber dynamics, the final
spin configuration depends on the spatial dimension: Lower dimensional systems tend to
arrive at the ground state more easily.
A one-dimensional system always reaches fully ferromagnetic phase
in the end \cite{Spirin01}.
Contrarily, in three-dimensional lattices, complex ferromagnetic domains
fluctuate forever,
which is called a blinker state \cite{Spirin01,Spirin01B,Sundaramurthy05,Olejarz11}.
As for complex networks such as the classical random network,
the small-world network, and the scale-free network
show similar behavior as three-dimensional cases
\cite{Haggstrom02,Boyer03,Castellano05,Castellano06,Herrero09}.
It can be understood by their small-world or ultra-small-world property,
which implies an infinite-dimensional system \cite{Daqing11}.
Another notable case is the complete graph, which arrives at the ground state always very fast \cite{Castellano05}.
In the complete graph, the shortest-path-length between nodes is always
one, and it can be regarded as a zero-dimensional system.

In two-dimensional lattices, which is between one-dimensional and three-dimensional cases,
the fate depends on the lattice type.
For the square and the triangular lattices, the Ising model reaches
the ferromagnetic ground state with probability of about 2/3 (for the aspect ratio of one),
and the other cases end in metastable states that have straight stripe patterns
\cite{Spirin01,Spirin01B,Sundaramurthy05,Lipowski99,Oliveira06,Barros09,Olejarz12,Blanchard13}.
As for the honeycomb lattice, it is always trapped in a metastable state \cite{Takano93,Blanchard17}.
It was argued that the fate is determined by whether the coordination number is even or odd \cite{Spirin01}.
However, the three lattice types are far from enough to make a conclusion
considering so many kinds of two-dimensional lattices.
In addition, the antiferromagnetic Ising model has not been studied in this context.
In the antiferromagnetic model, the geometric frustration, which exists in non-bipartite lattices,
changes the transition temperature and even the ground state \cite{Moessner06,Yu15,Yu16},
and it is expected to affect the fate of the zero-temperature Glauber dynamics, too.

In this paper, the fate of the ferromagnetic and antiferromagnetic Ising model
under zero-temperature Glauber dynamics from random spin configuration
on various two-dimensional lattices is studied.
We consider 31 Archimedean and 2-uniform lattices,
which are typically used for a systematic study in two dimensions \cite{Yu15,Yu16,Richter04,Neher08}.
We show that the classification by the coordination number is not always correct,
and a blinker state can be found also in two-dimension.

\section{Model and methods}
The energy of the Ising model is defined by the following Hamiltonian.
\begin{eqnarray}
H = -J \sum_{\langle i,j\rangle} S_i S_j
\label{Eq:Ising}
\end{eqnarray}
The spin at the $i$-th site $S_{i}$ may take the values of $+1$ (up) or $-1$ (down), only.
The summation $\langle i,j \rangle$ runs for all the nearest-neighbor spin pairs.
Positive and negative $J$ represent ferromagnetic and antiferromagnetic interactions, respectively.
The simulation starts from a random spin configuration,
which corresponds to the infinite temperature.
The beginning temperature is not important if it is  higher than the transition temperature.
The case where the system starts from the transition temperature is studied in \cite{Blanchard13}.
In the zero-temperature Glauber dynamics, a spin is randomly chosen
and the spin is flipped if the flipping reduce the energy.
When there is no change in energy by the spin flipping,
it is flipped with probability 1/2.
If a spin has the same direction as the majority of its neighbors,
the spin does not change its direction (type-I);
in the opposite case, the spin changes the direction when chosen (type-II);
if the two directions tie, the spin may be flipped or not (type-III).
Interestingly, the zero-temperature Glauber dynamics is exactly
the same as the majority-vote model without noise \cite{Oliveira92}.
The difference between the two models is the interpretation of the metastable state.
When all the spins are of type-I without fully ferromagnetic order,
it is a metastable state with higher energy than the ground state in the Ising model,
but all the spins (agents) are satisfied without frustration in the majority-vote model.

To accelerate the simulation, the continuous-time Monte-Carlo method \cite{Bortz75} was used.
Two lists are maintained for spins of type-II and type-III, respectively.
When the size of the two lists are $N_{\mathrm{II}}$ and $N_{\mathrm{III}}$,
one spin is chosen randomly with probability $2/(2N_{\rm II}+N_{\mathrm{III}})$
and $1/(2N_{\mathrm{II}}+N_{\mathrm{III}})$ in the lists of type-II and type-III, respectively.
The chosen spin is flipped and the lists are updated. By this method, fruitless
selection of type-I spin is avoided.
The simulation stops when there is no spin in the lists (ground state or metastable state)
or all the spins of type-III are blinker spins,
which flips forever, with $N_{\mathrm{II}}=0$ (blinker state).
The simulation is repeated $2\times 10^4$ times for each case and the
final states are classified into the ground state, metastable state,
and blinker state.

In this work, lattices made by edge-to-edge tiling by regular polygons are considered.
When a lattice has $k$ kinds of lattice
points with respect to the polygons surrounding the point,
it is called a $k$-uniform lattice. There are only eleven 1-uniform
lattices, also called the Archimedean lattices, and twenty 2-uniform
lattices \cite{Grunbaum87}, which are listed in Table~\ref{Table_fate} and Fig.~\ref{Fig_Arch}.
The periodic boundary condition is used. The shape of clusters is rectangle
with aspect ratio $r=l_y / l_x$, where $l_x$ and $l_y$ are the lengths
of the cluster in the $x$ and $y$ directions, respectively.

\section{Results and discussion} 

%%%%%%%%%%%%%%%%%%%%%%%%%%%%%%
\begin{figure}[tb!]
\includegraphics[width=13.0cm]{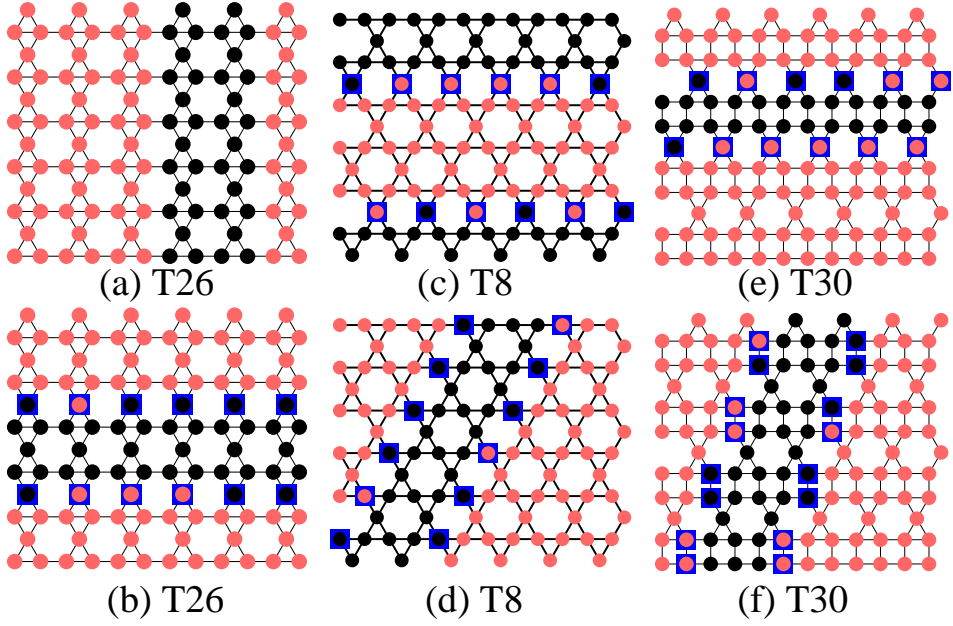}
\caption{Typical final spin configurations with stripes in the ferromagnetic Ising model.
Up-spins, down-spins, and blinker spins are expressed by black circles,
red circles, and blue squares, respectively.
}
\label{FIG_stripe}
\end{figure}

\begin{figure}[tb!]
\includegraphics[width=13.0cm]{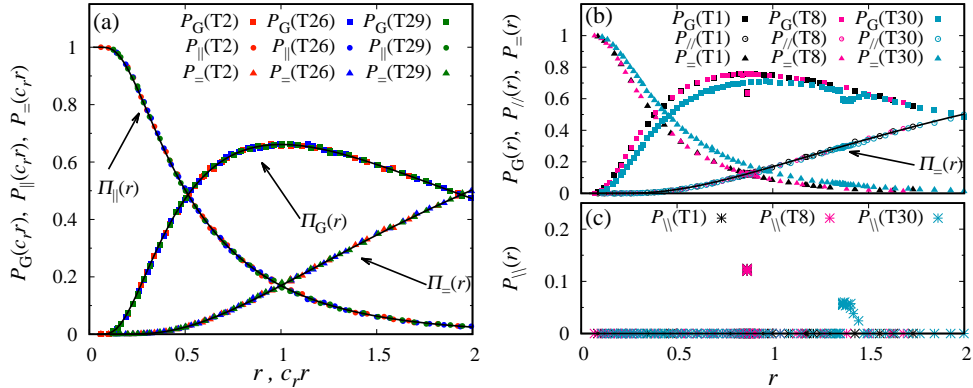}
\caption{The probabilities for final state of the zero-temperature Glauber dynamics as a function of the aspect ratio $r=l_y / l_x$.
$P_\mathrm{G}$, $P_{||}$, $P_{=}$, $P_{/\!/}$, and $P_{\setminus\!\setminus}$ mean the probabilities of the
ground state, and stripe states in the vertical, horizontal, and diagonal directions.
Rescaling factor $c_r$ was obtained by fitting to be $0.822(2)$ and $0.920(1)$ for T26 and T29, respectively.
For the other lattices, $c_r=1$ within statistical error.
Black solid lines represent the winding probability of the critical percolation
in the vertical and horizontal directions [$\mathit{\Pi}_{||}(r)$ and $\mathit{\Pi}_{=}(r)$]
and the other cases [$\mathit{\Pi}_{\mathrm{G}}(r) = 1-\mathit{\Pi}_{||}(r)-\mathit{\Pi}_{=}(r)$].
The number of spins ($N$) is between $600$ and $480000$.
The error bars are of similar size as the data points.
}
\label{FIG_r}
\end{figure}

\begin{figure}[tb!]
\includegraphics[width=13.0cm]{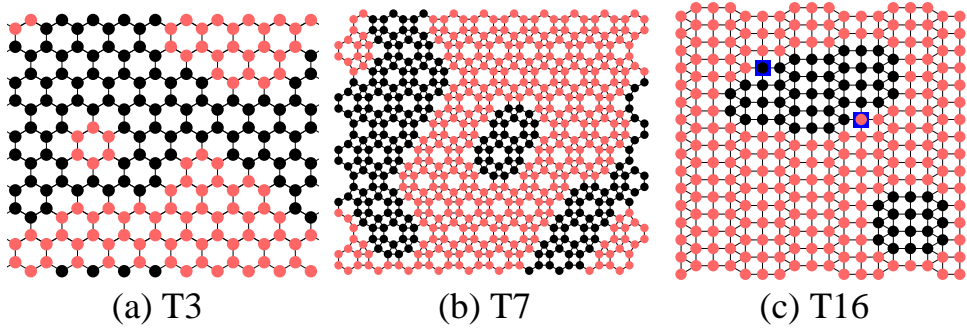}
\caption{Typical final spin configurations of a metastable state in (a) and (b),
and of a blinker state in (c).
Up-spins, down-spins, and blinker spins are expressed by black circles,
red circles, and blue squares, respectively.
}
\label{FIG_MB}
\end{figure}

\begin{figure}[tb!]
\includegraphics[width=13.0cm]{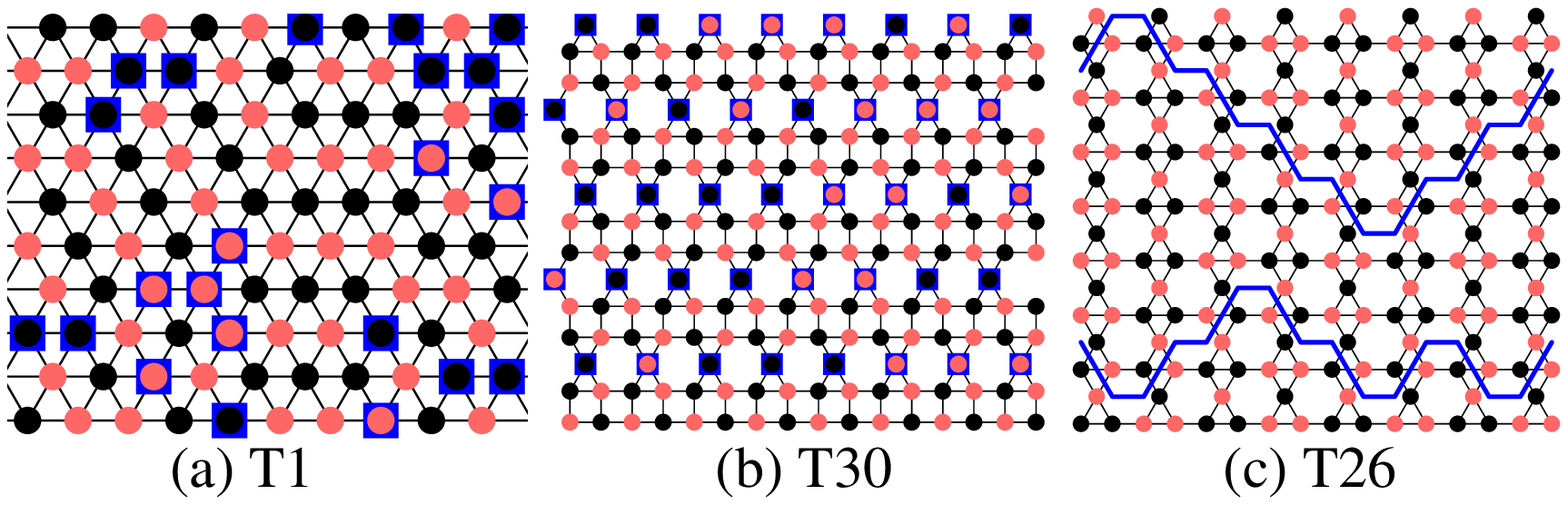}
\caption{Typical final spin configurations in the antiferromagnetic Ising model
for T1, T30, and T26.
Up-spins, down-spins, and blinker spins are expressed by black circles,
red circles, and blue squares, respectively.
In (c), domain boundary is represented by blue thick lines.
}
\label{FIG_AFM}
\end{figure}
%%%%%%%%%%%%%%%%%%%%%%%%%%%%%%

The fate of the 31 lattices considered here can be classified
into three groups.
Lattices of the first group arrive at the ground state 
or a stripe phase in the end.
Six lattices including the triangular (T1) and the square (T2) lattices
belong to this group.
The stripe phase is a metastable state for the triangular and the square lattices,
as is known already \cite{Spirin01}.
In T8 and T30, blinker spins separate the two domains in a stripe state, as shown in Fig.~\ref{FIG_stripe}.
Isolated blinker spins, which are surrounded by stationary spins unaffected by the sign of the
blinker spins, cannot move or expand, and fluctuate forever.
They can appear alone like in Fig.~\ref{FIG_stripe}(b)-(e) or
in pairs like in diagonal stripe state of T30 [Fig.~\ref{FIG_stripe}(f)]. 
Single blinker spins fluctuate between up and down spins, and blinker pairs in T30
wander among $(\uparrow\uparrow)$, $(\uparrow\downarrow)$, and $(\downarrow\downarrow)$ states.
In the case of T26 and T29, the final state can be a blinker state or a metastable state 
depending on the direction of the stripes.

Among the six lattices of the first group,
three lattices (T2, T26, and T29) may have stripe phase in the two directions (horizontal and vertical directions)
and the probabilities of them ($P_{=}$ and $P_{||}$) depend on the aspect ratio $r$.
[See Fig.~\ref{FIG_r}(a).]
As for the square lattice (T2), the dependence on the aspect ratio
was proposed to have relation to the critical percolation \cite{Barros09},
which is confirmed again in this work.
Soon after a zero-temperature quenching, the system approaches
the critical phase quickly \cite{Arenzon07,Sicilia07}, and a percolated cluster evolves to a stripe domain
when it exists. Therefore, the probability of the final stripe phase in a certain
direction is the same as the probability of percolation in the direction
at the critical percolation, which is approximated as follows \cite{Pruessner04,Barros09}.
\begin{eqnarray}
\mathit{\Pi}_{||}(r) = 1-2 \rho^{5/4} + \rho^{3}  + 2\rho^{12} - 4\rho^{5/43} + 3\rho^{15} \label{P_perc1}\\
\mathit{\Pi}_{=}(r) = \sqrt{\frac{2r}{3}} \left(\rho^{3} - \rho^{15}  - \rho^{27} + 4\rho^{35}\right) \label{P_perc2}\\
\mbox{with } \rho = e^{-\pi/6r} \label{P_perc3}
\end{eqnarray}
where $\mathit{\Pi}_{||}(r)$ ($\mathit{\Pi}_{=}(r)$) is the universal probability
of single or multiple separate windings in the vertical (horizontal) direction,
but not in the other
direction for a rectangle of aspect ratio $r$. The other kinds of domains
are unstable and so the probability to arrive at the ground state is
$\mathit{\Pi}_{\mathrm{G}}(r) = 1 - \mathit{\Pi}_{||}(r) - \mathit{\Pi}_{=}(r)$.
Equations~(\ref{P_perc1})-(\ref{P_perc3}) are accurate at least within 0.1\% \cite{Pruessner04}
and are good enough to compare with our results. Figure~\ref{FIG_r}(a) shows
that the agreement between the simulation results and the analytic results is
excellent. For T26 and T29, which have different property in the horizontal
and the vertical directions, a rescaling factor $c_r$ is needed to apply
the analytic results derived for the square lattice. They were obtained
by fitting of $\mathit{\Pi}_{\eta}(r) = P_{\eta}(c_r r)$
to be $c_r = 0.822(2)$ and $0.920(1)$ for T26 and T29, respectively.
The other three lattices (T1, T8, and T30) may have stripe states in three directions
in addition to the ground state. By the restriction of the periodic boundary
condition, however, stripes of only two directions are possible in general.
As shown in Fig.~\ref{FIG_r}(c), in the boundary condition used here,
stripe states of the third direction are allowed only at $r=\sqrt{3}/2$ for T1 and T8,
and near $r=(1+\sqrt{3})/2$ for T30. For other values of aspect ratio, the probability
of stripe states in that direction becomes zero;
this is reflected in many data of $P_{\setminus\!\setminus}(r)=0$
in Fig.~\ref{FIG_r}(c).
For lattices with diagonal stripe states, only the horizontal stripe phase
can be compared with the analytic solution of the critical percolation \cite{Blanchard13}.
As shown in Fig.~\ref{FIG_r}(a)-(c), the dependence on the cluster size is not observed
for $N \gtrsim 600$, where $N$ is the number of spins within the cluster.

Lattices of the second and the third groups result in
metastable state and blinker state, respectively.
The honeycomb lattice (T3) has a metastable final state,
as is known \cite{Takano93,Blanchard17}.
Typical metastable and blinker states as a fate are shown in Fig.~\ref{FIG_MB};
The boundary of domains can bend and an isolated domain island
can exist inside of a domain of the opposite spin.
Therefore, the possibility of the ground state
becomes zero as the cluster size increases. Depending on the details
of the lattice structure, a blinker spin can appear at the interface.
The probability of blinker spins also increases with the cluster size
if it is possible to appear. Stripe states can also appear with metastable
or blinker states. \cite{Blanchard17}

The classification of fate is summarized in Table~\ref{Table_fate}.
When a lattice includes odd-coordinated lattice points, it always belongs to
the second or the third group, and cannot arrive to the ground state.
Only even-coordinated lattices can belong to the first group.
However, even-coordinated lattices are not always in the first group:
we found a few even-coordinated lattices (T7, T17, T18, and T28)
that always fall into a metastable state or a blinker state, against common expectation.

We also investigated the fate of the antiferromagnetic Ising model on the 31 lattices
to study the frustration effect on the zero-temperature Glauber dynamics.
Bipartite lattices (T2, T3, T10, and T11), which do not have frustration, behave exactly the same way
as the ferromagnetic case;
in non-bipartite lattices, however, frustration tends to suppress the order \cite{Moessner06}.
While weak frustration just reduces the transition temperature,
strong frustration eliminates long-range order even at zero temperature resulting
in the spin ice or the spin liquid phase \cite{Yu15,Yu16}. 
Both of spin ice and spin liquid
do not have long-range order, but the former is frozen below freezing temperature while the latter
fluctuates forever even at zero temperature. Therefore, the spin liquid phase
is similar to the blinker state.
The triangular (T1) and the kagome (T8) lattices, which are spin liquid at the ground state
even by slow cooling \cite{Yu15}, become blinker state as is expected. [See Fig.~\ref{FIG_AFM}(a).]
A lattice that has a blinker final state in the ferromagnetic Ising model, always becomes a blinker state
in the antiferromagnetic case; A lattice of a metastable final state in the ferromagnetic case
may become a metastable state or a blinker state with antiferromagnetic interaction.
The results are summarized in Table~\ref{Table_fate}.

Most of non-bipartite lattices have a metastable or a blinker state without long-range ordering,
but there are two notable exceptions: T30 and T26. As shown in Fig.~\ref{FIG_AFM}(b),
T30 always has a stripe pattern in the final state: 
long-range-ordered two-leg ladders are separated by a string of blinker spins.
Since interactions between the ladders are blocked by blinker spins, each ladder can be considered
as one-dimensional effectively, and full long-range ordering is realized on the ladders. 
T26 is the only non-bipartite lattice that may have the complete
long-range-ordered ground state as a fate
in the antiferromagnetic Ising model among lattices considered here.
It may also have three kinds of stripe state in horizontal and two diagonal directions.
Vertical stripe states are unstable and always result in the ground state.
As shown in Fig.~\ref{FIG_AFM}(c), a horizontal stripe is not straight but in zigzags.
As for the aspect ratio of $\sqrt{3}/2$, where all the three stripe structures are possible,
$P_G=0.900(2)$, $P_{=}=0.069(1)$,  $P_{/\!/}=0.016(1)$,  $P_{\setminus\!\setminus}=0.015(1)$.
Size dependence was not noticed for $2700 \le N \le 874800$.

\section{Summary}

The fate of the Ising model by quenching through the zero-temperature Glauber dynamics is investigated
in the 31 two-dimensional lattices (Archimedean and 2-uniform lattices). They are classified into
three groups: (i) the ground state and stripe states, (ii) a metastable state, and (iii) a blinker state.
As for the ferromagnetic Ising model, contrary to common expectation that
even-coordinated and odd-coordinated lattices
belong to the group (i) and (ii), respectively,
a blinker state was found to exist in two dimensions.
In addition, we showed that even-coordinated lattice can 
belong to the group (ii) or (iii) to have a
metastable or a blinker state without stripe structure
as final state.
The universal relationship between the critical percolation and
the probability of striped final state
was confirmed for six lattices.
Study of the antiferromagnetic Ising model showed that frustration 
suppresses ordering more and promotes blinker state
rather than metastable and long-range ordered state.

\ack
%\section*{Acknowledgments}
This work was supported by GIST Research Institute (GRI) grant funded by the GIST in 2017.

%\References
\section*{References}
\bibliography{fate}
\end{document}